# Edge states and corner modes in second-order topological phononic crystal plates

Shao-yong Huo, Hong-bo Huang, Lu-yang Feng, Jiu-jiu Chen[*]

State Key Laboratory of Advanced Design and Manufacturing for Vehicle Body, College of Mechanical and Vehicle Engineering, Hunan University, Changsha 410082, People's Republic of China

[*]Email: jjchen@hnu.edu.cn

**Abstract:** We realize an elastic second-order topological insulator hosting both one-dimensional gapped edge states and zero-dimensional in-gap corner modes in the double-sided pillared phononic crystal plates with square lattice. Changing the width of two neighbor pillars breaks the inversion symmetry and induces the band inversion to emulate the quantum spin Hall effect where the gapless edge states are obtained. Further breaking the space-symmetry at interface, the gapless edge states are gapped and inducing the edge topological transitions and then giving rise to the zero-dimensional in-gap corner modes. Our work offers a novel way for elastic wave trapping and robustly guiding.

The discovery of the topological phases of matter has renewed our understanding of condensed matter physics due to its most fascinating property for the wave propagation with unique robustness to the defect and disorder. Driven by this discovery, a series of studies of topology are excavated from quantum electronics to classical bosonic systems, which include the photonic/phononic analogues of quantum Hall effect with broken time-reversal symmetry[1,2] and quantum spin Hall effect (QHSE) with time-reversal invariants,[3-5] Floquet topological insulators by temporal modulation,[6,7] topological valley states with broken spatial inversion symmetry.[8-12] Especially, with the rapid development of topological concepts, many revolutionary progresses have been made in technologies and applications, such as the superdirectional topological refraction antennas,[13,14] programmable coding insulator,[15,16] and on-chip integrated topological mechanical insulators.[17,18]

Recently, originated from the family of topological phases of matter, higher-order topological insulator[19-22] featuring with the corner states has drawn tremendous interest because it provides a new avenue to confine the waves, which obeys an extended topological bulk-boundary correspondence principle. For example, the two-dimensional (2D) second-order topological insulators (TIs) exhibit the one-dimensional (1D) gapped edge states and zero-dimensional (0D) gapless corner states. These multidimensional topological phase theories are first predicted for electronics[19,20] and then experimentally verified in topological mechanical[23] and electromagnetic metamaterials[24]. Inspired by them, the higher-order topological states are experimentally observed in acoustic kagome tight-blinding lattice in which the corner modes just locate at the acute-angle.[25,26] Meanwhile, another way to achieve the acoustic higher-order TIs was proposed by combining the crystalline symmetry of sonic crystals instead of the tight-blinding model where QSHE could be mimicked in square lattice.[27] The in-gap corner modes are formed based on the Jackiw-Rebbi soliton[28] mechanism, which shows the well confinement of second-order topological state at the right-angled corner.

The plate-mode wave propagation in phononic crystals (PnCs) has exhibited extraordinary behaviors, from complete band gap,[29] asymmetric transmission,[30] to negative refraction, which have promoted many exciting applications, such as elastic waveguides, diodes and sensors. Owing to the unique characteristic immune to the backscattering, the topological edge state of plate-mode wave has been extremely studied[11,12,17,18,31-34] for the potential in reinforcing nondestructive testing, high sensitivity sensing and information processing. However, due to the complicated mechanism of Bragg scattering, hybridization and local resonance in the PnCs plate structures[35,36], three modes polarizations (shear horizontal, symmetrical and anti-symmetrical Lamb modes) are intricately coupled, which make the realization of the elastic second-order TIs being still challenging, especially for that with topological complete bandgap.

In this paper, based on the Jackiw-Rebbi soliton mechanism, we utilize a model of double-sided pillared PnCs plate to mimic the QSHE in square lattice, in which the elastic 1D gapped edge states and 0D in-gap corner modes are obtained. A four-fold degeneracy can be constructed at *M* point of Brillouin zone (BZ) boundary by the band folding. The complete bandgaps are opened through changing the width of two neighbor pillars to break the inversion symmetry. The band topology exhibits a hierarchy of dimensions that the 1D edge states are determined by the bulk band phase transition and the 0D corner states are determined by the phase transition of edge band. We calculate the gapless edge state, gapped edge state and in-gap corner states in a topological complete bandgap which are further proved to be immune to the defects. In contrast with the previous reported second-order corner states in sonic crystals, the study of plate-mode wave may be greater potential in practical application, such as ultrasonic emission testing and elastic filters.

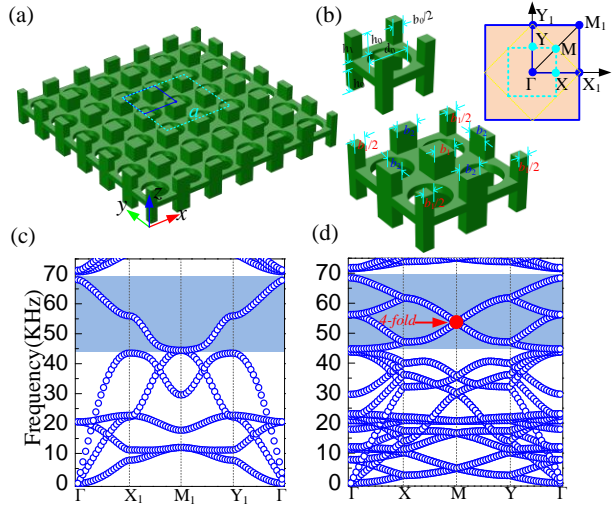

FIG. 1. (a) Schematic of square PnCs plate consisting of a perforated plate with double-side pillars. (b) The least unit cell, enlarged unit cell and the corresponding BZ, respectively. Band structures for least square unit cell (c) marked by blue solid-line in (a) and for enlarged square unit cell (d) marked by cyan dashed-line in (a).

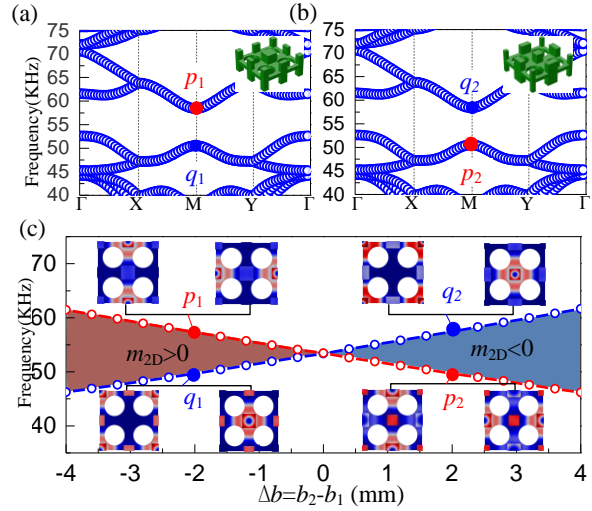

Fig. 2. Band structure induced by the broken inversion symmetry with $b_1=9$mm and $b_2=7$mm in (a) and $b_1=7$mm and $b_2=9$mm in (b). (c) Topological phase diagram of 2D plate-mode wave bands. The frequencies of odd-parity and that of even-parity at $M$ point are plotted by blue and red lines, respectively.

Figure 1(a) shows the considered model made up of a perforated PnCs plate stubbed with the double-side square pillars. The plate and pillars are made of the aluminum whose materials parameters are the Young's modulus $E_a$=70GPa, Poisson ratio $\sigma_a$=0.33 and mass density $\rho_a$=2700kg/m³. The thickness of perforated plate is $h_1$=3mm and the diameter of hole is $d_0$=14mm. The width and height of the pillars are $b_0$=8mm and $h_0$=8mm, respectively. The least unit cell and enlarged unit cell are displayed in Fig. 1(b). By using a finite element method, the band structure of the least unit cell is calculated as shown in Fig. 1(c). It can be seen that the sixth band is separate and non-crossed with other bands, which has been confirmed as a plate-mode band coupled by anti-symmetric Lamb mode and shear horizontal mode in the previous literature.[37] To construct the four-fold degeneracy, a strategy of band folding is employed by doubling the lattice along the $x$ and $y$ direction, which causes the shrinking of the BZ. The BZ boundary of the least unit cell corresponds to the blue solid line and that of enlarged unit cell ($a$=40mm) is narrowed to a half corresponding to the cyan dashed line. When taking the enlarged unit cell, a double Dirac dispersion occurs on the $M$ point displayed in Fig. 1(d), which is interpreted as the result of glide symmetries, $G_x := (x, y) \to (\frac{a}{2}+x, \frac{a}{2}-y)$ and $G_y := (x, y) \to (\frac{a}{2}-x, \frac{a}{2}+y)$. Combining the time-reversal operation $T$, the anti-unitary symmetry operators $\Theta_j = G_j * T (j = x, y)$ makes the four-fold degeneracy along the $MX$ and $MY$ boundary lines.[27]

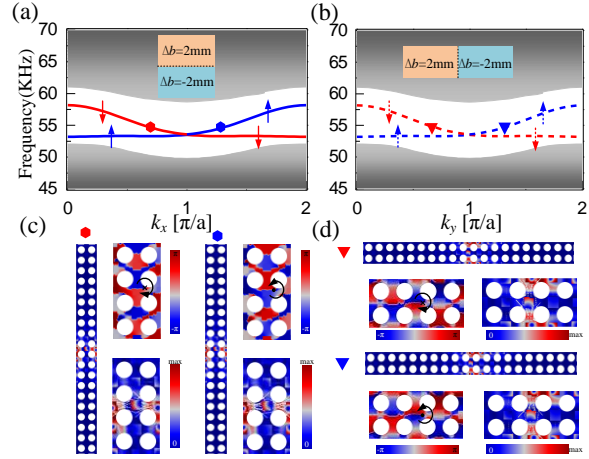

Fig. 3. Supercells band structures along the *x* (a) and *y* (b) directions between two PnCs plates with Δ*b*=2mm and Δ*b*=-2mm, respectively. The distributions of displacement fields, phases and mechanical energy flux of the edge states are displayed in (c) and (d) marked by two featured points, respectively.

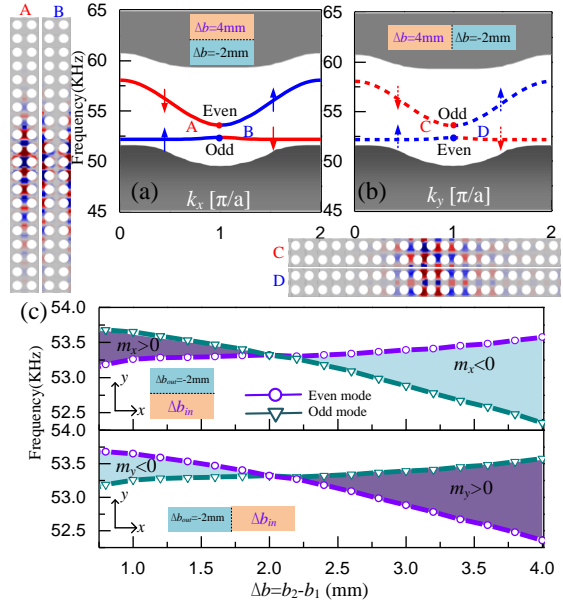

Fig. 4. Supercells band structures along *x* (a) and *y* (b) directions between two PnCs plates with Δ*b*=4mm and Δ*b*=-2mm, respectively. The displacement field distributions of gapped edge modes for the $k_x = \frac{\pi}{a}$ ($k_y = \frac{\pi}{a}$) point are presented by A and B (C and D). (c) Topological phase diagram for the edge states as plotted by the frequency of the even and odd modes at $k_x = \frac{\pi}{a}$ and $k_y = \frac{\pi}{a}$, respectively.

To split the four-fold degeneracy, we make the two neighbor pillars have different widths to break the inversion symmetry. For convenience, we define a parameter $\Delta b$ ($\Delta b=b_2-b_1$) to characterize the breaking of inversion symmetry. As shown in Fig. 2(a) and 2(b), when the width differences are $\Delta b$=-2mm and 2mm, respectively, the four-fold degeneracy is lift into two two-fold degeneracies with opposite parity distributions. For $\Delta b$=-2mm, a pair of degenerate odd-parity modes (marked by $p_1$ consisting of $p_x$- and $p_y$-like states) are distributed above the complete bandgap and a pair of degenerate even-parity modes (marked by $q_1$ consisting of $s$- and $d$-like states) are located below the complete bandgap. However, for $\Delta b$=2mm, two pairs of degenerate modes still locates on both side of the complete bandgap but their positions are flipped with degenerate even-parity modes (marked by $q_2$ consisting of $d$- and $s$-like states) on the upper side and degenerate odd-parity modes (marked by $p_2$ consisting of $p_y$- and $p_x$-like states) on the lower side. Furthermore, Fig.2(c) displays the bulk band topological phase transition process with the evolution of $\Delta b$ from a negative value to a positive one. It is worth noting that although this four-band model is built on the high-symmetry $M$ point, its parity inversion has the underlying physics similarity with acoustic pseudospins Hall insulators based on formed Hamiltonian theory from the Bernevig-Hughes-Zhang model[38] (seeing supplementary material). A Dirac mass $m_{2D}$ is used to characterize different topological phases whose signs are determined by the frequency difference between even-parity modes and odd-parity modes. When the frequency of odd-parity modes is higher than that of even-parity modes, the Dirac mass $m_{2D}$ is negative and the PnCs bandgap is endowed with topological nontrivial.

The featured manifestation of the analogical QSH insulator is the appearance of a pair of pseudospin-dependent edge states at interface with different topological properties. To confirm it, we construct a supercell which is stacked by two kinds of unit cells ($\Delta b$=2mm and $\Delta b$=-2mm) with different phases along the *x* direction and *y* direction, respectively. The projected bands are calculated as shown in Figs. 3(a) and (b). As expected, the plate-mode wave edge modes characterizing the topological helical states are found at the interface. Figures 3(c) and (d) present the distributions of the displacement field, phase and mechanical energy flux at two eigenmode points with opposite orbital angular momenta for edge states along *x*-direction and *y*-direction, respectively. It can be seen that the displacement fields are confined at the interfaces. The phase vortices also exhibit a characteristic of anticlockwise or clockwise distribution, which indicates the nature of the pseudospin-up or pseudospin-down, respectively, as displayed by blue and red arrows in Figs. 3(a) and (b). Seeing from Fig.3, two edge states emerge and they are degenerate at the $k_x = \frac{\pi}{a}$ ($k_y = \frac{\pi}{a}$), which is because the glide symmetry remains on the edges.

Next, we further consider the situation that the glide symmetry of edge is broken. A simple way is employed by making the supercell lattices on both sides of the interface have different modulus of $\Delta b$ simultaneously with different phases. As shown in Fig. 4(a), when the supercell is stacked by the unit cells of $\Delta b$=4mm and $\Delta b$=-2mm, the degenerate edge states are gapped and a bandgap is opened at the $k_x = \frac{\pi}{a}$ point along the $x$ direction which provides a chance to create the in-gap corner states. Importantly, two edge states host an even mode and an odd mode on the upper and lower sides of the edge bandgap, respectively. Furthermore, we calculate the supercell band structure along the $y$ direction and the degenerate edge states are also split while their positions are flipped with an odd mode distributed upper the edge bandgap and an even mode distributed lower the edge bandgap. Similar to the case of 2D, such gapped edge states can be characterized by 1D massive Dirac equation description of Hamiltonians (seeing supplementary material). Therefore, the 1D Dirac mass $m_x$ ($m_y$) at the $x(y)$ edge is introduced to quantify the edge bandgap and the $m_x$ ($m_y$) can be measured by the frequency difference of even and odd modes at $k_x = \frac{\pi}{a}$ ($k_y = \frac{\pi}{a}$) point because the mirror symmetry along the $x$ or $y$ direction still remains valid. To investigate the phase transition of edge states, we fix the lattices with $\Delta b$=-2mm at one end and adjust $\Delta b$ of lattices at other end as an input. From Fig. 4(c), we can see that the edge bandgap undergoes a topological transition from opening to closing and then to opening which is accompanied by the 1D edge Dirac mass ranging from a non-zero value to zero and then to non-zero. However, the 1D edge Dirac masses have opposite signs along the *x* and *y* directions. Such a sign difference will lead to the formation of Jackiw-Rebbi soliton modes at the junction of two edges.[27] To verify this point, we construct a box-shaped plate which is made up of two PnCs with the Δ*b*=-2mm wrapping up the Δ*b*=4mm as shown in the illustration of Fig. 5(a). We perform the eigenfrequency calculation for the combined PnCs plate where the continuously periodic boundary conditions are imposed along the upper and lower edges as well as the left and right edges, respectively. As expected, it can be seen that three kinds of states appear as shown in the Fig. 5(a) corresponding to the bulk state marked by the black color, edge states marked by olive color, and corner states marked by red and magenta colors, respectively. The corner states are almost degenerate at the frequency of 52.8 KHz. The displacement field distributions for the corner modes and edge states are exhibited in Fig. 5(b). Different from previous exploration in acoustic, additional four cracked corner states appear in the gap of edge states due to the greater impedance in solid plate system with complex boundaries. Finally, we further demonstrate that the obtained corner states have a good behavior of being against the disorders and defects (seeing supplementary material). Therefore, the bulk-edge correspondence have successfully been extended into the elastic plate system in a hierarchy of dimensions: the bulk band topology gives rise to the edge states and the edge states topology gives rise to the corner states. Such multidimensional topological transition rule is hopeful to realize the high-order topological states for bulk elastic waves.

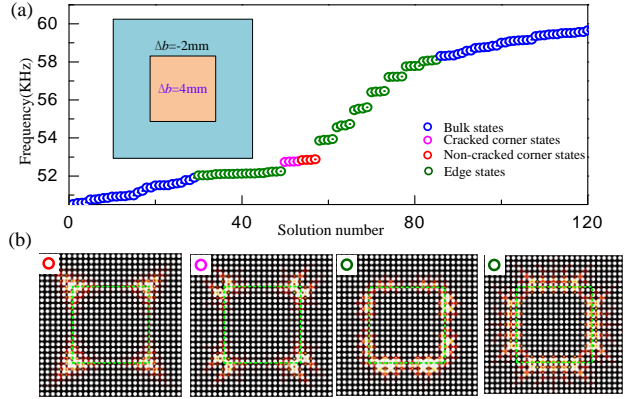

Fig. 5. (a) Calculated spectrum for a box-shaped structure consisting of an area of PnCs plate (Δb=4mm) with $7\times7a^2$ enlarged unit cells surrounded by lattices of Δb=-2mm with thickness $3a$ enlarged unit cells. (b) The displacement filed distributions of the corner states and edge states.

In conclusion, we have theoretically and numerically demonstrated an elastic second-order TI in pillared PnCs plates where the 1D gapped edge states and 0D in-gap corner states are achieved. The results show that the elastic topological helical edge states can be realized in square lattice by mimicking the QSHE at the high-symmetry *M* point. In addition, we have demonstrated that in the elastic plate, the 1D edge states topology leads to the 0D in-gap corner states. Moreover, the obtained edge states and corner states are located in topological complete bandgap. Finally, the corner states are proved to have the robustness immune to the defects and disorders. Our work may be an excellent candidate for practical applications including reinforcing nondestructive testing and enhancing energy harvesting.

**Acknowledgments**

The authors gratefully acknowledge financial support from National Science Foundation of China under Grant No.11374093 and Young Scholar fund sponsored by common university and college of the province in Hunan.